\documentstyle[aps,pre,epsf,floats]{revtex}

\begin{document}


\draft
\twocolumn[\hsize\textwidth\columnwidth\hsize\csname@twocolumnfalse%
\endcsname

\title{Why is the DNA Denaturation Transition First Order?}
\author{Yariv Kafri, David Mukamel and Luca Peliti\cite{Naples}\\[2mm]}

\address{Department of Physics of Complex Systems,
Weizmann Institute of Science, Rehovot 76100, Israel}
\date{\today}
\maketitle

\begin{abstract}
We study a model for the denaturation transition of DNA in which
the molecules are considered as composed of a sequence of
alternating bound segments and denaturated loops. We take into
account the excluded-volume interactions between denaturated loops
and the rest of the chain by exploiting recent results on scaling
properties of polymer networks of arbitrary topology. The phase
transition is found to be {\it first order} in $d=2$ dimensions
and above, in agreement with experiments and at variance with
previous theoretical results, in which only excluded-volume
interactions within denaturated loops were taken into account. Our
results agree with recent numerical simulations.
\end{abstract}

 ]

Thermal denaturation or melting of double-stranded DNA is the
process by which the two strands unbind upon heating. The nature
of this transition has been investigated for almost four decades
\cite{WB,PS1,Lifson,PB,CH,CM,TDP}. Experimentally, a sample
containing molecules of a specific length and sequence is
prepared. Then the fraction of bound base-pairs as a function of
temperature, referred to as the melting curve, is measured through
light absorption, typically at about 260~nm. For heterogeneous
DNA, where the sequence contains both AT and GC pairs, the melting
curve exhibits a multistep behavior consisting of plateaus with
different sizes separated by sharp jumps. These jumps have been
attributed to the unbinding of domains characterized by different
frequencies of AT and GC pairs. The sharpness of the jumps
suggests that the transition from bound to unbound is first order.

\begin{figure}
\begin{center}
\epsfxsize 8 cm
\epsfbox{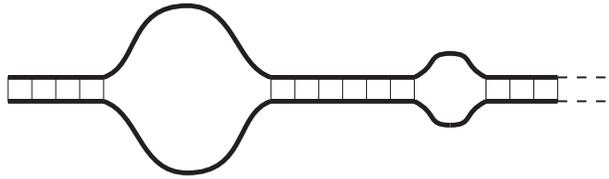}
\end{center}
\caption{Schematic representation of the Poland-Scheraga
model.}\label{model}
\end{figure}

The early theoretical models \cite{PS1,Lifson}, which we refer to
as Poland-Scheraga (PS) type models, consider the DNA molecule as
composed of an alternating sequence of bound and denaturated
states (see, e.g., Fig.~\ref{model}). Typically a bound state is
energetically favored over an unbound one, while a denaturated
segment (loop) is entropically favored over a bound one. Within
the PS type models the segments which compose the chain are
assumed to be non-interacting with one another. This assumption
considerably simplifies the theoretical treatment and enables one
to calculate the resulting free energy. In the past the entropy of
the denaturated loops has been evaluated by modelling them either
as ideal random walks~\cite{PS2} or as self-avoiding walks
\cite{Fisher}. It has been found that within this approach the
denaturation transition of DNA is continuous both in two and three
dimensions and it becomes first order only above four dimensions.
It was suggested \cite{Fisher} that taking into account the
interaction between loops would sharpen the transition.

A version of the PS model which includes the excluded-volume
interactions between the various segments of the chain has been
recently introduced~\cite{CCG}. This extension makes the model
analytically intractable, but numerical studies in three
dimensions indicate that these interactions indeed drive the
transition first order.

In this Letter we consider analytically the effects of
excluded-volume interaction between the various segments of the
chain. Although we treat this interaction only in an approximate
way, we are able to give some insight into the unbinding mechanism
and on the nature of the transition. Our approach makes use of
recent important results on the entropy of self-avoiding polymer
networks \cite{Dup,SFLD}. We find that this interaction drives the
transition {\it first order} in $d=2$, 3 and $4-\varepsilon$
dimensions. To proceed we introduce and discuss the PS model. A
scaling argument which takes into account the interaction between
the segments of the DNA chain is then presented and used to study
the transition. Finally, we comment on disorder and compare our
results to those obtained from more recent models of DNA
denaturation \cite{PB,CH,CM,TDP}.

The PS model considers two strands, made of monomers, each
representing one persistence length of a single strand (ordinarily
$\sim 40$~\AA\  \cite{RMMRSU}). For simplicity we take boundary
conditions where the monomers at the ends of the molecule are
always bound. All other monomers on the chain can be either bound
or unbound to a specific matching monomer on the second chain. The
binding energy $E_0<0$ is taken to be the same for all matching
monomers. A typical DNA configuration is shown in
Fig.~\ref{model}. It is made of sequences of bound monomers
separated by denaturated loops. The statistical weight of a bound
sequence of length $k$ is $\omega^k = \exp(-kE_0/T)$, where $T$ is
the temperature~\cite{note}. On the other hand the statistical
weight of a denaturated sequence of length $k$ is given by the
change in entropy due to the added configurations arising from a
loop of length $2k$. For large $k$ this has the general form $A
s^k/k^c$, where $s$ is a non-universal constant and the exponent
$c$ is determined by the properties of the loop configurations.
For simplicity, we set $A=1$. The model is most easily studied
within the grand canonical ensemble where the total chain length
$L$ is allowed to fluctuate. The grand canonical partition
function, ${\cal Z}$, is given by
\begin{equation}
{\cal Z}=
\sum_{M=0}^{\infty}G(M)z^M=\frac{V_0(z)U_L(z)}{1-U(z)V(z)},
\label{partition}
\end{equation}
where $G(M)$ is the canonical partition function of a chain of
length $M$, $z$ is the fugacity and
\begin{equation}
U(z)=\sum_{k=1}^{\infty}\frac{s^k}{k^c}z^k, \qquad
V(z)=\sum_{k=1}^{\infty}\omega^k z^k,  \label{UV}
\end{equation}
$V_0(z)=1+V(z)$ and $U_L(z)=1+U(z)$. Equation (\ref{partition})
can be verified by expanding the partition function as a series in
$U(z)V(z)$. The factors $V_0(z)$ and $U_L(z)$ properly account for
the boundaries. The average chain length, $\langle L \rangle$, is
set by choosing a fugacity such that $\langle L \rangle=\partial
\ln {\cal Z}/\partial \ln z$. We wish to evaluate the order
parameter $\theta$, i.e., the fraction of bounded pairs, as a
function of temperature, in the limit $\langle L \rangle \to
\infty$. The average number of bounded pairs in a chain is given
by $\langle m \rangle= \partial \ln {\cal Z}/\partial \ln \omega$,
so that
\begin{equation}
\theta= \lim_{L \to \infty} \frac{\langle m \rangle}{\langle L
\rangle} =\frac{
\partial \ln z^* }{\partial \ln \omega}. \label{theta}
\end{equation}
Here $z^*$ is the value of the fugacity in the limit $\langle L
\rangle \to \infty$. This is the lowest value of the fugacity for
which the partition function (\ref{partition}) diverges, i.e., for
which $U(z^*)V(z^*)=1$. Using $V(z)=\omega z /(1- \omega z)$ one
has
\begin{equation}
U(z^*)=1/(\omega z^*)-1. \label{transtion}
\end{equation}

It is clear that the nature of the denaturation transition is
determined by the dependence of $z^*$ on $\omega$. The function
$U(z)$ is independent of $\omega$. It is finite for $z<1/s$ and
diverges when $z>1/s$. On the other hand the function $1/V(z)$
increases continuously as $\omega$ decreases (corresponding to an
increase in $T$). Thus, as the temperature increases from $T=0$,
the fraction of attached monomers decreases and $z^*$ increases.
However if $z^*$ reaches the value $z_{\rm c}=1/s$ (so that
$1/V(1/s)\ge U(1/s)$) any further increase of the temperature
leaves $z^*$ unchanged so that $\theta=0$. Therefore the
transition takes place at $z^*=z_{\rm c}=1/s$. Its nature is
determined by the behaviour of $U(z)$ in the vicinity of $z_{\rm
c}$. This is controlled in turn by the value of the exponent $c$.
There are three regimes:
\begin{enumerate}
\item For $c\le 1$, $U(z_c)$ diverges, so that $z^*$ is an
analytic function of $\omega$ and no phase transition takes place.
\item For $1<c\le 2$, $U(z_c)$ converges but its derivative
$U'(z)$ diverges at $z^*=z_c$. Thus $\theta \propto \partial z^* /
\partial \omega$ vanishes continuously
($\theta \sim |T-T_{\rm c}|^{(2-c)/(c-1)}$) at the transition and
the transition is {\it continuous}.
\item For $c>2$, both $U(z)$ and its derivative converge at
$z^*=z_c$ and the transition is {\it first order}. Here, in
contrast to the continuous case, the average size of a denaturated
loop is finite at the transition.
\end{enumerate}
The value of the exponent $c$ can be obtained by enumerating
random walks which return to the origin, so that $c=d \nu$. For
ideal random walks this yields $c=d/2$: thus there is no
transition at $d\le 2$, a continuous transition for $2<d \le 4$
and a first order transition only for $d>4$ \cite{PS2}. On the
other hand, for self-avoiding random walks the excluded volume
interaction modifies the exponent to $c=3/2$ for $d=2$ and $c
\approx 1.764$ for $d=3$. The transition is thus sharper, but
still continuous, in three dimensions~\cite{Fisher}.

In the above treatment the entropy of a loop of length $2l$ was
taken to be of the form $s^l/l^c$. In determining the exponent $c$
the interaction of the loop with the rest of the chain has been
ignored. In the following we argue that, when the interaction of
the loop with the rest of the chain is taken into account, the
above form for the loop entropy is still valid, but the exponent
$c$ is modified. The excluded volume interaction of the loop with
the rest of the chain is expected to lower the entropy of the loop
yielding a larger value of the exponent $c$.

We evaluate the exponent $c$ in this situation by considering the
number of configurations of a loop of length $2l$ embedded in a
chain of length $2L$ (see Fig. 2). In reality the chain itself is
composed of an alternating sequence of loops and bound segments of
various sizes. However, in the present analysis we ignore the
detailed structure of the rest of the chain and consider it as
being in a bound state. The fact that the rest of the chain is
made of both bound and unbound segments increases the excluded
volume interaction and is thus expected to increase further the
value of $c$. This point will be discussed later.

In order to evaluate the exponent $c$ for our configuration, we
use the results obtained by Duplantier \cite{Dup,SFLD} for the
entropy of general polymer networks: Consider a polymer network
composed of segments and vertices with an arbitrary given
topology. The results state that for a network composed of
segments with an average length $L_{\rm A}$ the number of
configurations is proportional to $\propto s^{L_{\rm A}} L_{\rm
A}^{\gamma_{\rm G}-1}$ where $\gamma_{\rm G}$ is an exponent which
depends {\it only} on the topology of the network. It is given by
\begin{equation}
\gamma_{\rm G}=1 - d \nu  {\cal L} + \sum_{N\geq 1} n_N \sigma_N,
\label{Dup1}
\end{equation}
where ${\cal L}$ is the number of independent loops, $n_N$ is the
number of vertices with $N$ outgoing legs, and $\sigma_N$ is an
exponent related to such a vertex. The exponents $\sigma_N$ are
exactly known in $d=2$ and in $d=4-\varepsilon$ to
$O(\varepsilon^2)$.

Consider now the relevant topology to the problem discussed in
this work (Fig.~2). We are interested in enumerating the
configurations of the network in the limit $l/L \ll 1$, when the
loop size is much smaller than the rest of the chain. To do this
we note that when only two length scales determine the geometry of
the network, as in the case depicted in Fig.~2, the number of
configurations can be written as~\cite{SFLD}
\begin{equation}
\Gamma = s^{L+l} (L+l)^{\gamma_{\rm loop}-1} g(l/L),
\label{scaling}
\end{equation}
for large $L$ and $l$. Here $g(x)$ is a scaling function and
$\gamma_{\rm loop}$ can be evaluated using eq.~(\ref{Dup1}). For
the topology considered above of a loop embedded in two segments
(Fig.~2) we have: one loop, ${\cal L}=1$; two vertices of order
one, $n_1=2$, corresponding to the two free ends of the chain
(denoted by $V1$ in the figure); two vertices of order three,
$n_3=2$ (denoted in the figure by $V3$). Using eq.~(\ref{Dup1}) we
obtain
\begin{equation}
\gamma_{\rm loop}=1- d \nu + 2 \sigma_1 + 2 \sigma_3.
\label{gloop}
\end{equation}
\begin{figure}
\begin{center}
\epsfxsize 8 cm
\epsfbox{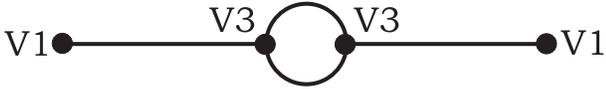}
\end{center}
\caption{The topology of the loop embedded in a chain. The
distance along the chain from a vertex of type $V1$ to the nearest
vertex of type $V3$ is given by $L$. The distance between the
vertices of type $V3$ is equal to $l$. }\label{top}
\end{figure}

We are interested in a loop size much smaller than the length of
the chain, $l/L \ll 1$. Clearly, in the limit $l/L \to 0$, the
number of configurations should reduce to that of a single
self-avoiding open chain, which, to leading order in $L$, is given
by $s^{L} L^{\gamma-1}$, where $\gamma=1+ 2\sigma_1$. This implies
that, in the limit $x \ll 1$,
\begin{equation}
g(x) \sim x^{\gamma_{\rm loop}-\gamma}. \label{limit}
\end{equation}
Thus the $l$-dependence of $\Gamma$, which gives the change in the
number of configurations available to the loop, is
\begin{equation}
\Gamma \propto s^l l^{\gamma_{\rm loop}-\gamma} \cdot s^L
L^{\gamma-1}. \label{gdep}
\end{equation}
It is therefore evident that for large $l$ and $L$ and in the
limit $l/L \ll 1$ the partition sum is decomposed into a product
of the partition sums of the loop and that of the rest of the
chain \cite{Note2}. The excluded volume effect between these two
parts reflects itself in the value of the effective exponent $c$
associated with the loop entropy. This result is very helpful
since it enables one to extend the PS approach to the case of
interacting loops. From Eq. \ref{gdep} one sees that the
appropriate effective exponent $c$ is
\begin{mathletters}
\label{loopc:eq}
\begin{equation}
c = \gamma-\gamma_{\rm loop}=d\nu-2\sigma_3\;. \label{cdef}
\end{equation}
In $d=2$ $\sigma_3=-29/64$ \cite{Dup} and $\nu=3/4$ yielding
\begin{equation}
c= 2+13/32 \;. \label{2d}
\end{equation}
In $d=4-\varepsilon$ to $O(\varepsilon^2)$ one has
$\sigma_3=-3\varepsilon/16+9\varepsilon^2/512$ and
$\nu=1/2(1+\varepsilon/8+15/4(\varepsilon/8)^2)$ yielding
\begin{equation}
c= 2+\varepsilon/8+ 5 \varepsilon^2/256\;. \label{4d}
\end{equation}
\end{mathletters}
In $d=3$ one may use Pad\'{e} and Pad\'{e}-Borel approximations to
obtain $\sigma_3 \approx 0.175$ \cite{SFLD} which with the value
$\nu \approx 0.588$~\cite{SFLD} yielding $c \approx 2.115$.

Equations~(\ref{loopc:eq}) are our main result. The fact that the
exponent $c$ is found to be larger than $2$ in $d=2,3$ and
$4-\varepsilon$ indicates that the transition is first order in $d
\ge 2$.

As noted above this approach does not take into account all the
excluded volume interactions in the system since the rest of the
chain is assumed to be in a bound state. In practice however the
rest of the chain is by itself composed of loops and bound
segments of various sizes. The existence of loops is expected to
enhance the excluded volume interaction. Taking into account all
these configurations in the partition sum is a formidable task.
However, we estimate below the effect of these interactions on the
partition sum and show that it does not modify the main conclusion
of this analysis, namely that the transition is first order.
\begin{figure}
\begin{center}
\epsfxsize 7 cm \epsfbox{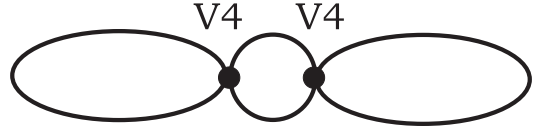}
\end{center}
\caption{An extreme topology where the loop of length $2l$ is
between two denaturated loops of size $2L$ each. The vertices of
order $4$ are denoted by $V4$.}\label{top2}
\end{figure}

For a given loop-bound configuration of the rest of the chain it
is straightforward to check that the $l$ dependence of the
partition sum (\ref{gdep}) in the scaling limit is valid as long
as the two bound segments which are attached to the loop under
consideration are long. On the other hand when at least one of
these segments is short and the adjacent open loop is large the
effective exponent $c$ is modified. In order to estimate this
exponent we consider the extreme case where the rest of the chain
is composed of two loops each of size $2L$ (see Fig. \ref{top2}).
A similar analysis to that presented above yields for the value of
$c$,
\begin{mathletters}
\label{loops}
\begin{eqnarray}
c &=& d\nu- \sigma_4 \;, \\ &=& 2 + 11/16\;,
\;\;\;\;\;\;\;\;\;\;\;\;\;\;\;\;\;\;\;\; {\rm in} \;\; d=2 \;,
\\ &=& 2+\varepsilon/4- 15 \varepsilon^2/128 \;, \;\;\;\;\; {\rm in}
\;\; d=4-\varepsilon,
\end{eqnarray}
\end{mathletters}
where the values $\sigma_4=-19/16$ in $d=2$ and
$\sigma_4=-\varepsilon/2+11(\varepsilon^2/8)^2$ in
$d=4-\varepsilon$ dimensions~\cite{Dup} are used along with those
of $\nu$. Using $\sigma_4 \approx 0.46$ obtained by Pad\'{e} and
Pad\'{e}-Borel approximations gives in $d=3$ the value $c \approx
2.22$. It is easy to verify that this scaling behavior is
unchanged as long as one of the bound states connected to the loop
under consideration is small (so there is at least one vertex of
order four). Evidently $c$ for this configuration is larger than
that of the configuration in Fig. 2. One can show that for
arbitrary loop-segment configuration of the chain the effective
$c$ is given by either Eqs. (\ref{loopc:eq}) or Eqs. (\ref{loops})
depending on the length of the bound segments attached to the
loop. Since in both cases the exponent $c$ is found to be larger
than $2$ in $d=2$, $d=4-\varepsilon$ and in $d=3$, this analysis
strongly suggests that the transition is first order for $d \ge
2$. The analysis presented above is valid as long as the length
$l$ of an individual loop is much smaller than the total length of
the chain. The fact that the transition is found to be first
order, so that the average loop size remains finite at the
transition, makes the analysis self consistent.

The model treated above assumes, for simplicity, that the monomers
at the ends of the two chains are bounded. In reality the ends
need not be attached and one should consider open boundary
conditions. Similar analysis for the case where one of the ends of
the chain is open yields the same result. Note that one end of the
chains must remain attached for a bound state to exist. In the
above approach entangled configurations with non equivalent knot
topologies, which are possible in $d=3$, are counted \cite{Dup} as
they should when one end of the chains is open.

Finally, we compare the model to recently studied models for the
denaturation transition. These have focused on the stacking energy
generating a varying stiffness in the bound and denaturated state
\cite{PB,CH,TDP} or on the helical structure of the DNA molecule
\cite{CM}. All these studies are within a directed polymer
approach and thus ignore the effects studied in this Letter.

In summary, we have shown that the reduction in the number of
available configurations of denaturated loops due to the
excluded-volume interactions between a denaturated loop and the
rest of the chain is sufficient to drive the denaturation
transition first order. This is done by extending the PS type
models to include these interactions. Our results have a direct
implication on recent numerical simulations of the
model~\cite{CCG}. The model studied here treats the binding energy
between matching monomers is taken to be equal, although each base
can bind only to its corresponding base on the other strand. This
last assumption is reasonable when the heterogeneity of the chain
is taken into account. For short segments to bind the detailed
sequences of the two segments have to match and this is less
likely to take place in a heterogeneous system. It is therefore
important to consider the effect of the heterogeneity of the
binding energy between the matching base pairs on the order of the
transition. Also, another effect which has recently been
emphasized is the varying stiffness of the phosphate backbone. The
model considered above does not treat these effects.

It is instructive to note that although the average loop size is
finite at the transition, the loop size distribution is rather
broad and decreases as $1/{l^c}$ for large $l$. For $c<3$ this
yields a diverging variance of the loop size. It would be of great
interest to test this prediction experimentally.

We are indebt to M. J. E. Richardson and H. Orland for many
inspiring discussions and their involvement earlier stages of our
study of the DNA denaturation transition. We also thank B.
Duplantier for very helpful discussions and M. E. Fisher for
useful comments on the manuscript. LP wishes to pay a tribute to
Raffaele (Peo) Tecce, who introduced him to the beauties of
biological systems. The support of the Israeli Science Foundation
is gratefully acknowledged.



\begin{thebibliography}{99}
\bibitem{note}
For simplicity, the configurational entropy related to the
embedding of the chain of bound monomers in ambient space is
neglected. One may check that it will not modify our main results.
\end{thebibliography}
\end{document}